\documentclass[twocolumn,showpacs,preprintnumbers,amsmath,amssymb,floatfix]{revtex4}
 
\usepackage{graphicx}
\usepackage{dcolumn}
\usepackage{bm}
\usepackage{dsfont}
\begin{document}  

\title{The QCD phase diagram according to the center group}

\author{Ydalia Delgado Mercado$^a$, Hans Gerd Evertz$^b$, Christof
Gattringer$^a$
\vspace{2mm}}

\affiliation{$^a$Institut f\"ur Physik, Karl-Franzens 
Universit\"at, Graz, Austria \\
$^b$Institute for Theoretical and Computational Physics, Technische Universit\"at
Graz, Austria}

\begin{abstract}
We study an effective theory for QCD at finite temperature and density which
contains the leading center symmetric and center symmetry breaking terms. The
effective theory is studied in a flux representation where the complex phase
problem is absent and the model becomes accessible to Monte Carlo techniques  
also at finite chemical potential. We simulate the system using a generalized
Prokof'ev-Svistunov worm algorithm and compare the results to a low temperature expansion. The phase
diagram is determined as a function of temperature, chemical potential and quark
mass. Shape and quark mass dependence of the phase boundaries are as
expected for QCD. The transition into the deconfined phase is smooth throughout,
without any discontinuities or critical points.
\end{abstract}
\maketitle

\noindent
Obtaining a deeper understanding of the QCD phase diagram will be one of the central
goals of particle physics in the coming years. With running and upcoming
experiments that drive this development,  also the theoretical side is challenged to
improve our understanding of the QCD phase structure. Analyzing phase transitions is
clearly a non-perturbative  problem and suitable techniques have to be applied. For
vanishing chemical potential, lattice QCD is a powerful method that provides
reliable quantitative information on the QCD finite temperature transition. However,
for non-vanishing density the notorious complex phase problem so far limits
numerical lattice QCD studies to painfully small volumes.

For quenched QCD, where the quark contributions to the path integral are neglected,
the deconfinement transition is related  to the center group  $\mathds{Z}_3$ of
SU(3), which is a symmetry in the confined low temperature phase, while it is broken
spontaneously above the deconfinement temperature \cite{znbreaking}. When one
couples the dynamics of the quark fields, the center symmetry is broken explicitly
by the fermion determinant. This explicit breaking overlays the spontaneous breaking
of the quenched theory. However, as for spin systems, one may expect that  also for
QCD the underlying symmetry still governs parts of the dynamics of the full theory,
e.g., via the center properties of canonical determinants \cite{canonical}. 

In order to study the role of center symmetry for the QCD phase diagram, we here
analyze an effective theory which contains the leading center symmetric and center
symmetry breaking terms. This \emph{effective center theory} can be mapped exactly
to a flux representation \cite{flux}, where the complex phase problem is absent. 
So far the model was studied only in a very limited parameter range 
\cite{flux,Alford,Kim,Forcrand,Condella}.  
Here we apply a generalization of the worm
algorithm \cite{worm} which allows us to efficiently 
explore the full range of temperatures and chemical
potential values. This constitutes one  of the very few examples where a QCD related
complex phase problem can be solved. We study the phase diagram of the  effective
center theory and analyze what role the center degrees of freedom of QCD play for
the phase structure of hot and dense matter.

\vskip4mm
\noindent
{\bf Effective center theory and flux representation}
\vskip1mm
\noindent
The effective center theory is defined by the action \cite{flux}
\begin{equation}
S[P]  = - \!\sum_x \left(\! \tau \! 
\sum_{\nu = 1}^3 \!
\Big[ P_x P_{x+\hat{\nu}}^* + c.c. \Big] + 
\eta P_x +  \overline{\eta} P_x^*\! \right) .
\label{action_original}
\end{equation}
The dynamical degrees of freedom are the center variables $P_x \in
\mathds{Z}_3  =  \{1,e^{i 2 \pi/3}, e^{-i 2\pi/3}\}$ at the sites $x$  of a
3-dimensional cubic lattice.  
The partition function is a sum over all 
configurations of the center variables, $Z= \sum_{\{P\}}\exp(-S[P])$.  The first
term of the action (\ref{action_original}) is a nearest neighbor  interaction which
is invariant under global center transformations ($P_x \rightarrow z P_x$), 
where all  variables are
transformed  with a center element $z \in \mathds{Z}_3$.
The form of this term may be obtained from a strong coupling expansion of the
effective action for the Polaykov loop which in quenched QCD is the order
parameter for center symmetry and for confinement.  The variables $P_x$ take
over the role of the local Polyakov loops. Although in full QCD center symmetry
is broken explicitly by the quarks, the Polyakov loop is still used to monitor
confinement properties and to determine the crossover temperature (see, e.g.,
\cite{fodor,hotqcd}). The strong coupling expansion also identifies the parameter 
$\tau$ as an increasing function  of the temperature $T$ of the underlying
lattice QCD theory, and for brevity we refer to $\tau$ as \emph{temperature}. 

The second term of the effective action may be obtained from a hopping  (i.e., large
quark mass) expansion of the fermion determinant and contains the leading center
symmetry breaking contributions. The parameters $\eta  =  \kappa e^{\, \mu} \, , \;
\overline{\eta} =  \kappa e^{-\mu}$  are related to the chemical potential $\mu$.
The hopping expansion shows that $\kappa = N_f \, h(m)$, where $N_f$ is the number 
of flavors 
and $h(m)$ is a function of the QCD quark mass $m$ which decreases with increasing $m$. 
We refer to $\kappa$ as the
\emph{inverse mass parameter}.  

For vanishing external field, $\kappa = 0$, 
the model reduces to the 3-state Potts model, which is known to have a first order 
transition at $\tau = 0.183522(3)$ \cite{potts}. For small external field
$\kappa$ and vanishing $\mu$  
the first order transition persists, giving rise to a short first order line
which terminates at a critical endpoint at $(\tau,\kappa) = (0.183127(7),0.00026(3))$
\cite{potts}. For small non-zero $\mu$ the system has been analyzed
with techniques based on the Swendsen-Wang cluster algorithm \cite{Alford},
with reweighting \cite{Kim} and with imaginary $\mu$  
\cite{Forcrand}. Within the flux representation \cite{flux} local Metropolis 
updates were also used \cite{flux,Condella,Alford}. It has  been demonstrated that
turning on the chemical potential softens the transition and shifts the 
critical endpoint towards smaller values of $\kappa$. 
So far no simulations were
done in the parameter region where the complex phase problem of the formulation
(\ref{action_original}) becomes severe (see \cite{Alford} for a discussion of
that regime). 

The flux representation \cite{flux} solves the complex phase problem. We 
briefly summarize it to discuss our observables and conventions: 
The Boltzmann factors for the
nearest neighbor terms of (\ref{action_original}) can be rewritten as 
\begin{equation}
e^{\, \tau [ P_x P_{x+\hat{\nu}}^* + c.c.]} \, =  \, C \!\!\!\!
\sum_{b_{x,\nu} = \, -1}^{+1} \!
B^{|b_{x,\nu}|} \;\, ( \, P_x P_{x + \hat{\nu}}^{\,*} \, )^{b_{x,\nu}} \; .
\label{linkfactor}
\end{equation}
The sum on the right hand side is over \emph{flux variables} 
$b_{x,\nu} \in \{-1,0,+1\}$ attached to the links of the lattice. The
constants $C$ and $B$ depend on the temperature $\tau$ via  
$C =(e^{2\tau} + 2 e^{-\tau})/3$ and   
$B = \; (e^{2\tau} - e^{-\tau})/3C$. 
Similarly one expresses the center symmetry breaking terms as
\begin{equation}
e^{\, \eta \, P_x \; + \; \overline{\eta} \, P_x^*}  \; = \;
\sum_{s_x=-1}^{+1} M_{s_x} \, P_x^{\, s_x} \; ,
\label{sitefactor}
\end{equation}
where we sum over \emph{monomer variables} $s_x \in \{-1,0,+1\}$
attached to the sites $x$. It is straightforward to work out the monomer 
weights $M_s$ for $s = -1,0,+1$,
\begin{equation}
\hspace*{-1mm}M_s = \frac{1}{3} \! \Big[ e^{ \eta + \overline{\eta}}  +  
2 e^{-(\eta + \overline{\eta})/2} 
\cos\Big(\frac{\sqrt{3}}{2}(\eta\!-\!\overline{\eta}) - s
\frac{2\pi}{3}\Big)\! \Big] .\!
\label{monoweights}
\end{equation}
The weights $M_s$ turn out to be non-negative.

Inserting (\ref{linkfactor}) and (\ref{sitefactor}) into the partition sum 
gives rise to a complete factorization of the dependence on the dynamical 
variables $P_x$ and the sum over all configurations can be performed. One ends
up with a new form for the partition sum 
(we drop an irrelevant overall constant),
\begin{equation}
Z = \!\! \sum_{\{b,s\}} \! 
W(b,s) \, \prod_x  T \Big( \sum_\nu [b_{x,\nu} - b_{x-\hat{\nu},\nu}] 
+ s_x \Big).
\label{zflux}
\end{equation}
The partition function is now a sum over configurations 
$\{b,s\}$ of the flux and monomer 
variables. Each configuration comes with a real non-negative weight factor
$W(b,s) =  ( \prod_{x,\nu} B^{|b_{x,\nu}|} )\, ( 
\prod_x \, M_{s_x} ).$
For every link with non-vanishing flux, i.e., $b_{x,\nu} = \pm 1$,
a factor $B$ is taken into account. Sites $x$ 
contribute with factors $M_{s_x}$ according to their monomer values 
$s_x \in \{-1,0,+1\}$.

The flux/monomer configurations are subject to constraints: In (\ref{zflux})
$T(n)$ is the triality
function defined as $T(n) = \delta_{\,n \, \mbox{\small mod} \, 3 \, , \, 0}$.
The constraints enforce that at every site $x$ the total  flux from both,
flux variables $b_{x,\nu}$ and monomers $s_x$, is a multiple of 3. 
In the flux form (\ref{zflux}) the partition sum 
contains only real and non-negative contributions and thus the complex phase
problem is solved. 

In this letter we focus on bulk observables such as the
order parameter $P$ and the corresponding susceptibility, 
which both  are obtained
as derivatives of the free energy, $\langle P \rangle = \partial \ln Z /
\partial \eta$ and $\chi_P = \partial^2 \ln Z /
\partial \eta^2$. In a similar way one  obtains the internal energy $U$
and the
heat capacity $C$. For an efficient evaluation the identities
$\partial M_{+1}/\partial \eta = M_0$, $\partial M_{0}/\partial \eta = M_{-1}$
and $\partial M_{-1}/\partial \eta = M_{+1}$ are useful. In the end all
our observables are expressed in terms of the total
flux and monomer numbers 
and their moments. 
      
\vskip4mm
\noindent
{\bf Simulation with the worm algorithm}
\vskip1mm
\noindent
Having established the flux representation where the complex phase problem is
solved, we now must find a suitable algorithm for an efficient Monte Carlo 
update. We here use a generalized form of the Prokof'ev-Svistunov worm 
algorithm \cite{worm}: The worm starts at a randomly chosen site and 
moves along links until it returns to the starting point where it terminates. 
We allow for two different moves of our worm:
A) The worm randomly chooses a new direction at a site and changes the flux at the
corresponding link by $\pm 1$. 
B) The worm decides to change a monomer variable by $\pm 1$
and then randomly hops to another site where the monomer variable is changed 
by $\mp 1$.
The moves are offered with equal probability, produce only configurations that 
are compatible with the constraint, and lead to an ergodic algorithm. The Metropolis 
acceptance probabilities are $p_A = B^{\delta_b}$ when changing 
a flux variable $b$ by an amount of $\delta_b$ (Move A), and
$p_B = M_{s^\prime}/M_s$ for changing a monomer variable from $s$ to $s^\prime$ (Move
B). A more complete account of the algorithm
and its implementation will be given elsewhere.  

We generated ensembles for lattice sizes $36^3$, $48^3$, $64^3$ and $72^3$. 
For the inverse mass parameter we used $\kappa = 0.1, \kappa = 0.01,
\kappa = 0.005$ and $\kappa = 0.001$. The evaluation of our observables 
$\langle P \rangle$, $\chi_P$, $U$ and $C$ is based
on up to $10^6$ configurations, separated by 10 worms for
decorrelation. Autocorrelation times were determined  
and used in the estimate for the statistical errors. Finite volume effects were
analzed by comparing the different system sizes and are negligible for
our final results.  

The results from the new worm algorithm were checked using several strategies. For
vanishing inverse mass parameter the known results \cite{potts} for the 3-state
Potts model with external magnetic field were reproduced. For small values of $\tau$
and arbitrary $\kappa$ and $\mu$ we used low temperature expansion techniques to
determine the power series for  the partition sum, taking into account all terms up
to $\tau^3$. For small $\tau$ we found excellent agreement between the Monte Carlo
results and the perturbative series (see Fig.~\ref{phasediagram_chiP}).
Finally, for all our production and analysis codes two independent programs were
written for cross checks.  
 
\begin{figure}[t]
\includegraphics[width=8.4cm,clip]{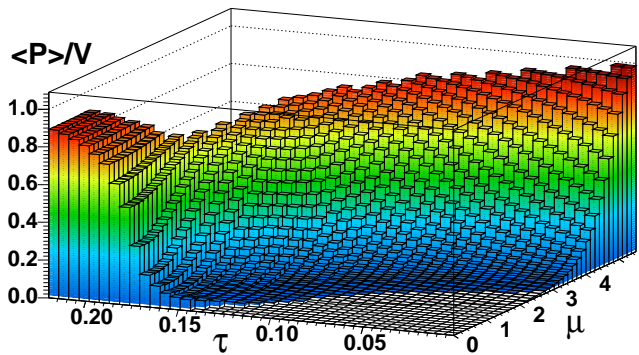}
\vspace*{-3mm}
\caption{The order parameter $\langle P \rangle / V$ as a function of temperature
$\tau$ and chemical potential $\mu$ from our $36^3$, $\kappa =
0.01$ ensembles. Near the rear left corner no data were computed. 
$\langle P \rangle / V$ has values close to 1 there.  For small
$\tau$ and $\mu$ the order parameter approaches 0.
\vspace*{-3mm}}
\label{orderparameter}
\end{figure}

\vskip4mm
\noindent
{\bf Results from the Monte Carlo calculation}
\vskip1mm
\noindent
We begin the discussion of our results with the order parameter $\langle P
\rangle$, which -- as discussed above -- is identified with the
Polyakov loop of QCD.
In Fig.~\ref{orderparameter} we show the results for $\langle P \rangle / V$ as
a function of $\tau$ and $\mu$ for our $36^3$ ensembles at $\kappa = 0.01$. For
roughly 450 points in the $\tau$-$\mu$ plane the values of 
$\langle P \rangle / V$
were evaluated and used for the 3-D plot. In the rear left corner and for small
$\tau$ and $\mu$ no data were
computed. $\langle P \rangle / V$ is expected to be close to 1 in the rear left
corner. For small $\tau$ and $\mu$ there is
a sizable region where the expectation value $\langle P \rangle / V$  is  
small and center symmetry is broken only very mildly. Transferring this finding
from the effective center theory to QCD implies that for small
temperature and density, matter is confined. 
When $\tau$ or $\mu$ are increased, the system undergoes a change and   
$\langle P \rangle / V$   
reaches values close to 1. For QCD this implies that both temperature
and $\mu$ may be used to drive the system into the deconfined  
phase characterized by a large Polyakov loop.

The next step in the analysis of the phase diagram is to  identify the phase
boundary. For that purpose we studied the susceptibility $\chi_P$ and the  heat
capacity $C$ as a function of $\mu$ at fixed $\tau$ (symbols with horizontal error
bars in Figs.~2 and 3) or as a function of $\tau$ at fixed $\mu$ (vertical error
bars) and determined the position of the maximum: We fitted
the data for $\chi_p$ and $C$ near the maxima with a parabola and obtained the
position of the maximum as one of the fit parameters. The corresponding statistical
error was computed with the jackknife method. In Fig.~\ref{phasediagram_chiP} we
show the positions of the maxima of $\chi_P$ in the $\tau$-$\mu$ plane. We compare
the results for 4 values of the inverse mass parameter $\kappa$ and connect the data
at the same $\kappa$ with a dotted line to guide the eye. The dashed horizontal line
at the top marks the value of the critical $\tau$ for the 3-state Potts model, i.e.,
the situation at $\kappa = 0$. The dashed curves near the bottom of the plot are the
results from the perturbative series for small $\tau$ which we briefly  discussed in
the last section. The Monte Carlo data nicely approach these curves for $\tau
\rightarrow 0$.

\begin{figure}[t!]
\hspace*{-1mm}
\includegraphics[width=8.1cm,clip]{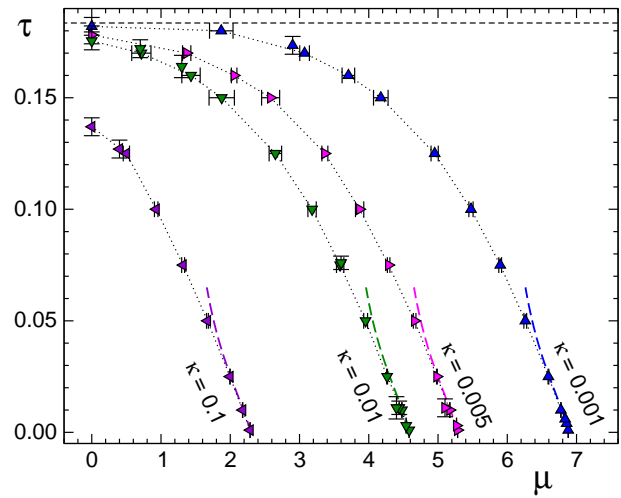} 
\vspace*{-3mm}
\caption{Phase diagram as obtained from the maxima of the Polyakov loop
susceptibility $\chi_P$. We show results at 4 values of the inverse mass
parameter $\kappa$. The dashed curves at the bottom are the results of
the $\tau$ expansion and the horizontal line marks the critical value of
$\tau$ for the $\kappa = 0$ case.\vspace*{-3mm}}
\label{phasediagram_chiP}
\end{figure}

The curves in Fig.~\ref{phasediagram_chiP} separate the phases with small 
order parameter and with $\langle P \rangle / V \sim 1$, i.e., the confined and
the deconfined phases. The phase boundaries depend on the inverse mass parameter 
$\kappa$, and their
behavior is as expected for QCD: The intercept of the phase lines with the $\mu$-axis
shifts to the left with decreasing quark mass (i.e., increasing inverse mass
parameter $\kappa$) because a smaller chemical potential is sufficient to excite
lighter states. Also the intercept with the $\tau$-axis drops with increasing
$\kappa$, corresponding to the fact that for quenched QCD (i.e., infinite
quark mass) the transition
temperature is considerably higher than the crossover temperature of QCD with
physical quark masses. The mass dependence of the phase boundaries thus
is as expected for QCD.

A key problem of the QCD phase diagram is the question
about the nature of the various transitions and phases. Unless one goes to very high
densities where more exotic phases exist, two principal phases are
expected. A phase with conventional matter (confined with broken chiral
symmetry) and a plasma phase (deconfined and chirally symmetric). In some
parameter regions also a quarkyonic phase with confinement but restored chiral
symmetry has been discussed. For the transition lines a standard
scenario is that at $\mu = 0$ for physical quark masses 
the finite temperature transition is merely a crossover
\cite{wuppnature}, where
different second derivatives of the free energy peak at different temperature
values \cite{fodor,hotqcd}. With increasing $\mu$ the crossover region narrows and
terminates at a critical endpoint. From the endpoint on a first order 
transition line continues (which at some point might hit other
transitions to the exotic phases mentioned). Alternative scenarios 
 suggest
that either no critical endpoint appears and crossover type of behavior 
persists also for large $\mu$ and low temperature, or that even more than one endpoint 
might exist. For a glimpse of the current debate see e.g.\ \cite{phasediagram}.

\begin{figure}[t!]
\hspace*{-1mm}
\includegraphics[width=8.0cm,clip]{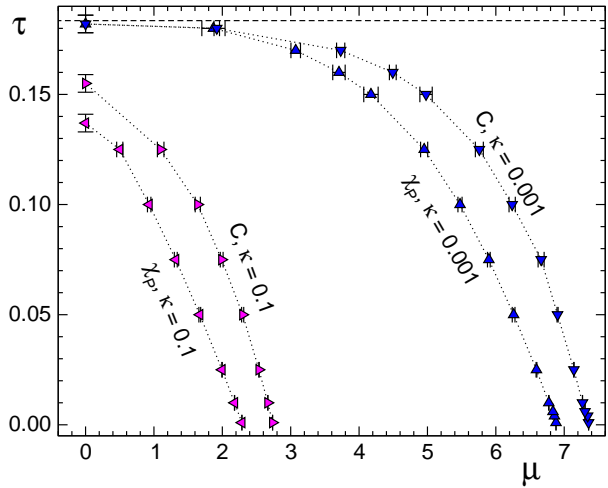}
\vspace*{-3mm}
\caption{Comparison of the phase boundaries obtained from the maxima of 
susceptibility 
$\chi_P$ and heat capacity $C$.  }
\label{phasediagram_chiP_C}
\vspace*{-5mm}
\end{figure}

While we cannot address questions concerning chiral symmetry in the effective
center theory, we can  analyze the type of transitions that take place
at the phase boundaries. To determine the nature of the transitions we used
two techniques. We analyzed histograms for the distribution of the order
parameter and the action near the phase boundaries. For a first order
transition the histograms would display a double peak structure near the
critical line. In our analysis we found only single peaks and thus rule out
first order behavior. The second approach is a comparison of the 
normalized susceptibilities and heat capacities $\chi_P / V$ and $C/V$ from
lattices with different volumes. For a first order transition the height of the
maxima diverges proportional to $V$, while for a continuous transition the
divergence is modified by a critical exponent. A height which is independent 
of $V$ indicates a smooth crossover. Our analysis shows that
for the phase boundaries of Figs.~\ref{phasediagram_chiP} and 3
the height is independent of $V$ for all volumes we studied. We conclude that 
the transitions in the
effective center theory are smooth crossover lines.  

Once the crossover nature is established, one may ask how wide the transition
region is -- similar to the finite temperature crossover of QCD at zero density,
which is 20 - 30 MeV wide. In order to get an estimate for the width of the crossover region, in
Fig.~\ref{phasediagram_chiP_C}  we compare the positions of the maxima of the
susceptibility $\chi_P$ and of  the heat capacity $C$ for two values of
$\kappa$. The fact that the corresponding lines do not coincide stresses again the crossover nature of the
transition, and the plot demonstrates that the crossover region is rather wide for most
of the parameter values. Only for small $\kappa$ and $\mu$ the lines approach each other,
anticipating the first order behavior which is known for very small $\kappa$ and $\mu$
\cite{Alford,Kim}.

\vskip4mm
\noindent
{\bf Concluding remarks}
\vskip1mm
\noindent
In this letter we report on our results for the phase diagram of an effective
theory for the center degrees of freedom of QCD. Using the flux representation
solves the complex phase problem and we develop a
generalized worm algorithm for a Monte Carlo calculation in a wide range of
temperatures and chemical potential. 

The outcome of our analysis is a version of the QCD phase diagram when only the
center degrees of freedom are considered. The phase diagram shares many
features with the conjectured full QCD phase diagram: The transition to the 
deconfined phase can be driven by both, temperature or $\mu$, and the quark 
mass dependence is as expected for QCD. The phase boundaries between a phase
with only very small center symmetry breaking ($\langle P \rangle / V \sim 0$)
and a phase with $\langle P \rangle / V \sim 1$
has a shape which is similar to the one
conjectured for QCD. For all parameter values we studied, the
transition is of a smooth crossover type and we conclude that
center symmetry alone does not
provide a mechanism for possible first order behavior in the QCD phase diagram. 

Various future research directions may be followed: The effective theory can
be made more realistic by replacing the $\mathds{Z}_3$ spins by continuous SU(3)
valued variables (here some work is in progress and also for this theory
the complex phase problem can be solved by a suitable flux representation). Furthermore
it would be desirable to take into account also the fermion nature of the
problem -- an aspect which is absent in the current effective action. Another
interesting direction is of a more technical nature: With our effective theory
we have a reference example of a QCD related system where the complex action
problem is solved. This reference theory can and should be used to test the
reliability and limitations of various techniques for QCD with chemical potential,
such as reweighting, series expansions or complex Langevin methods.

\vspace{-7mm}
\begin{acknowledgments}
\vspace{-3mm}
The authors thank P.~de Forcrand, C.B.~Lang, R.~Ritter, 
B.-J.~Schaefer and U.-J.~Wiese for valuable discussions and remarks. The project 
has been supported by FWF DK 1203 and Marie Curie ITN STRONGnet.
\end{acknowledgments}

\end{document}